\documentclass[
 reprint,
superscriptaddress,
amsmath,amssymb,
 aps,
pra]{revtex4-2}
\usepackage{xcolor}
\usepackage{graphicx}% Include figure files
\usepackage{dcolumn}% Align table columns on decimal point
\usepackage{bm}% bold math
\newcounter{stepcnt}          % 定义步骤专用计数器
\renewcommand{\thestepcnt}{\arabic{stepcnt}} % 编号用阿拉伯数字

\usepackage{hyperref}

\hypersetup{
	colorlinks=true,
	linkcolor=blue,
	citecolor=blue,
	urlcolor=blue,
	breaklinks=true
}

\begin{document}

\preprint{APS/123-QED}

\title{A self-tallying quantum anonymous voting protocol for multiple-selection elections}

\author{Guang-Bao  \surname{Xu}}
   \affiliation{College of Computer Science and Engineering, Shandong University of Science and Technology, Qingdao, 266590, China}
   \affiliation{Shandong Key Laboratory of Smart Mine Information Technology,  Qingdao, 266590, China}
\author{Meng-Na  \surname{Li}}
   \affiliation{College of Computer Science and Engineering, Shandong University of Science and Technology, Qingdao, 266590, China}
 \author{Yu-Guang \surname{Yang}}%
   \affiliation{College of Cyberspace Science and Technology, Beijing University of Technology, Beijing, 100124, China}
 \author{Dong-Huan \surname{Jiang}}
     \email{donghuan\_jiang@163.com}
   \affiliation{College of Mathematics and Systems Science, Shandong University of Science and Technology, Qingdao, 266590, China}

% \collaboration{MUSO Collaboration}%\noaffiliation

% \author{Charlie Author}
%  \homepage{http://www.Second.institution.edu/~Charlie.Author}
% \affiliation{
%  Second institution and/or address\\
%  This line break forced% with \\
% }%
% \affiliation{
%  Third institution, the second for Charlie Author
% }%
% \author{Delta Author}
% \affiliation{%
%  Authors' institution and/or address\\
%  This line break forced with \textbackslash\textbackslash
% }%

% \collaboration{CLEO Collaboration}%\noaffiliation

\date{\today}% It is always \today, today,
             %  but any date may be explicitly specified

\begin{abstract}
Self-tallying quantum anonymous voting (SQAV) has drawn considerable attention since it was proposed. However, constrained by design challenges, all existing protocols only accommodate single-selection voting and cannot support multi-selection voting. In this paper, we propose a SQAV protocol with multi-selection voting functionality. In our protocol, $nm$ $n$-particle entangled states are equally divided into $n$ groups, and the $n$ particles in each entangled state are delivered to $n$ voters, one particle per voter. Each voter obtains $n$ voting vectors by sequentially measuring all his (or her) particles that belong to $n$ different groups, and then encodes his or her voting information into the vector corresponding to the secret index. Due to entanglement correlation, each voter can verify whether his or her ballot is correct through computation, while any participant can obtain the vote count for each candidate through computation. Our protocol satisfies self-tallying, nonreusability, verifiability, and fairness. 

%\begin{description}
%\item[Usage]
%Secondary publications and information retrieval purposes.
%\item[Structure]
%You may use the \texttt{description} environment to structure your abstract;
%use the optional argument of the \verb+\item+ command to give the category of each item. 
%\end{description}

\end{abstract}

%\keywords{Suggested keywords}%Use showkeys class option if keyword
                              %display desired
\maketitle

%\tableofcontents

\section{INTRODUCTION}
\label{sec:Introduction}

Voting is widely used in collective decision-making, including political elections, organizational governance, and committee decisions. In electronic voting systems, cryptographic techniques are commonly used to protect ballot secrecy, voter privacy, and the integrity of the voting process \cite{Jonker2013Privacy}. Most classical electronic voting protocols rely on public-key cryptography based on computational complexity, such as integer factorization and the discrete logarithm problem \cite{Kho2022Review}. However, with the rapid development of quantum algorithms \cite{Shor1994Algorithms}, those cryptography protocols are facing growing challenges. Thus, the security of those classical electronic voting protocols, which are based on public-key cryptography, can hardly be guaranteed. Fortunately, quantum voting protocols, which are based on two fundamental principles of quantum mechanics \cite{Scarani2009Security}, i.e., the Heisenberg uncertainty principle and the no-cloning theorem \cite{Wootters1982Single}, can effectively resist attacks \cite{Bennett1984Quantum} from quantum algorithms .

Because it can preserve voter privacy, quantum anonymous voting (QAV) has attracted considerable research interest. Early studies on QAV mainly focused on hiding the link between a voter and his (or her) vote. Hillery et al. investigated communication privacy based on quantum resources and discussed potential generalizations for constructing voting protocols \cite{Hillery2006Privacy}. Vaccaro et al. subsequently proposed quantum protocols for anonymous voting and surveying. In their protocol, distributed entangled states are used to preserve vote anonymity \cite{vaccaro2007quantum}. Horoshko and Kilin proposed a quantum anonymous voting protocol with an anonymity check to protect voter privacy against a curious tallyman \cite{horoshko2011quantum}. Jiang et al. proposed continuous-variable QAV protocols for two-valued and multivalued ballots and introduced constraints to prevent voters from voting more than once \cite{jiang2012quantum}. Later studies further explored different quantum resources and voting architectures. For example, Liu et al. developed a QAV protocol based on high-dimensional single particles, in which a private position index allows each voter to trace their own ballot in an anonymous manner \cite{Liu2021Quantum}. These studies gradually extended QAV from basic anonymous vote casting to protocols with enhanced privacy, verifiability, and nonreusability properties.

In parallel with these security-oriented developments, another line of research has focused on extending the functionality of quantum voting beyond single-option or binary-option ballots. Li and Zeng proposed a quantum anonymous voting protocol for elections with multiple candidates \cite{Li2008Quantum}. More recently, Ruan et al. proposed a distributed QAV protocol based on discrete-modulated coherent states and an optical frequency comb \cite{Ruan2022Multiparty}. The protocol explicitly considers both single-selection and multiple-selection ballot scenarios and supports multiparty and multivalued voting. Zhao and Jiang also developed a traveling QAV protocol based on GHZ states, in which voters can choose from multiple ballot items rather than being restricted to binary ``yes'' or ``no'' decisions \cite{Zhao2023Novel}. These studies demonstrate that quantum voting can accommodate richer election rules. However, tallying in these protocols relies on designated parties or authorities, and self-tallying cannot be realized.

Self-tallying provides a different direction for the development of QAV. In a self-tallying election, the final result can be independently obtained from publicly available voting data without relying on a dedicated tallying authority. In 2016, Wang et al. introduced the first self-tallying quantum anonymous voting (SQAV) protocol based on two types of multipartite entangled states \cite{Wang2016SelfTallying}. The protocol simultaneously provides privacy, self-tallying, nonreusability, verifiability, and fairness. More recently, Yang et al. proposed a resource-efficient SQAV protocol based on two-particle entangled states \cite{Yang2025ResourceEfficient}. Compared with the original SQAV protocol, their protocol substantially reduces the quantum-resource requirement and further supports positive voting, negative voting, and abstention. Despite considerable progress, these existing self-tallying quantum anonymous voting (SQAV) protocols are limited to single-selection voting and fail to effectively address multi-selection scenarios. In Wang et al.’s protocol, the nonreusability constraint limits each voter to selecting only one candidate. Although Yang et al.'s protocol enhances quantum-resource efficiency and broadens the range of admissible voting options, it cannot extend SQAV to multiple-selection ballots. Therefore, the design of a SQAV voting protocol with multiple-selection capability is of great research significance. 

In this paper, we propose a SQAV protocol for multiple-selection elections. Our SQAV protocol is designed on the basis of $n$-particle entangled states, and offers anonymity, self-tallying, and multiple-selection capability. That is, our quantum voting protocol effectively addresses the problem of enabling each voter to select multiple candidates at once. This paper is structured as follows. In Sec.~\ref{sec:Preliminaries}, we introduce some preliminaries which will be used in our protocol. A detailed description of our SQAV protocol for multiple-selection elections is provided in Sec.~\ref{sec:SQAV}. We analyze the correctness and security of the proposed protocol in Sec.~\ref{sec:correctness-security-analysis}, including self-tallying, nonreusability, verifiability, and fairness. Finally, we conclude the paper and discuss future work in Sec.~\ref{sec:conclusion}.

\section{PRELIMINARIES}
\label{sec:Preliminaries}

Consider an \(n\)-level quantum system with the computational basis $\mathcal{B}_C=\{|k\rangle_C,\,k=0,\,1,\,\ldots,\,n-1\}$. The $n$-level discrete Fourier transform \(F_n\) maps the computational basis to the Fourier basis $\mathcal{B}_F=\{|j\rangle_F,j=0,1,\ldots,n-1\}$, where $|j\rangle_F=F_n|k\rangle_C=\frac{1}{\sqrt{n}}\sum_{k=0}^{n-1}\omega^{jk}|k\rangle_C$ snd \(\omega=e^{2\pi \sqrt{-1}/n}\).

The following $n$-particle, $n$-level entangled state is used as the basic quantum resource of the proposed protocol:
\begin{equation}
\lvert\Psi_n\rangle=\frac{1}{\sqrt{n}}\sum_{k=0}^{n-1}\lvert k,k\oplus1,\ldots,k\oplus(n-1)\rangle_C,
\end{equation}
where \(\oplus\) denotes addition modulo \(n\). In the Fourier basis, this state can be defined as
\begin{widetext}
\begin{equation}
\label{eq:psi_fourier}
|\Psi_n\rangle = \frac{1}{n^{(n-1)/2}}
\sum_{\substack{j_0,\ldots,j_{n-1}\in\mathbb{Z}_n \\
\sum_{r=0}^{n-1} j_r \equiv 0 \pmod{n}}}
\omega^{-\sum_{r=0}^{n-1} rj_r}
|j_0,j_1,\ldots,j_{n-1}\rangle_F.
\end{equation}
\end{widetext}

From these two representations, \(\lvert\Psi_n\rangle\) has the following two measurement properties. 

(1) When each particle of \(\lvert\Psi_n\rangle\) is measured in the computational basis \(\mathcal{B}_C\), the joint outcome is \((k,k\oplus1,\ldots,k\oplus(n-1))\) with probability \(1/n\) for some \(k\in\mathbb{Z}_n\). Thus, the measurement outcomes form a cyclic permutation of \(\mathbb{Z}_n\). 

(2) As shown in Appendix~\ref{app:fourier}, when each particle of \(\lvert\Psi_n\rangle\) is measured in the Fourier basis, only the joint outcomes satisfying \(\sum_{\ell=0}^{n-1}j_\ell\equiv0\pmod n\) have nonzero probability.

These two measurement properties are used in the subsequent protocol for secret-index generation, ballot-number generation, and security checks.

\section{QUANTUM ANONYMOUS VOTING PROTOCOL}
\label{sec:SQAV}

\subsection{Protocol Procedure}

Assume that the voting system consists of \(n\) eligible voters, denoted by \(V_0,\,V_1,\,\ldots,\,V_{n-1}\), and \(m\) candidates, denoted by \(C_0,\,C_1,\,\ldots,\,C_{m-1}\). The ballot of voter \(V_i\) is represented by a binary vector 

\begin{equation}
 \nonumber
 v_{i}=
\begin{bmatrix}
v_{0,i} \\
v_{1,i} \\
\vdots \\
v_{m-1,i}
\end{bmatrix},
\end{equation}
where
\[
v_{r,i}=\begin{cases}1, & \text{if voter } V_i \text{ casts a vote in favor of }  C_r\\0, & \text{if voter } V_i \text{ does not cast a vote for } C_r\end{cases}
\]
for $r=0,\,1,\,\ldots,\,m-1.$
If the voting rule allows each voter to select at most \(L\) candidates, a valid ballot must also satisfy \(\sum_{r=0}^{m-1} v_{r,i} \leq L\). The proposed protocol employs a trusted third party T, who is exclusively responsible for the preparation of entangled states and the distribution of particles, and does not take part in the subsequent measurements, ballot encoding, or vote tallying. All classical communication takes place over pairwise authenticated channels. During the data-publish phase, a simultaneous broadcast mechanism~\cite{hevia2005simultaneous} is employed, so that no participant can fix his or her disclosed values on the basis of the values already released by other participants. The protocol proceeds in six steps, as detailed below.

\smallskip
\noindent\hspace{1em}\refstepcounter{stepcnt}\textbf{Step \thestepcnt. Quantum-state preparation and distribution.}
\label{step1}
The trusted third party T prepares \(1+nm+n\delta\) identical entangled states \(|\Psi_n\rangle\), where \(\delta\) is a security parameter controlling the number of states used for eavesdropping detection. For each \(|\Psi_n\rangle\), T distributes its $n$ particles randomly among the $n$ voters, such that each voter obtains exactly one particle.

\smallskip
\noindent\hspace{1em}\refstepcounter{stepcnt}\textbf{Step \thestepcnt. Eavesdropping detection.}
\label{step2}
To detect eavesdropping, all the voters \(V_i\) randomly select \(\delta\) entangled states \(|\Psi_n\rangle\) as check states. For each check state, all the voters \(V_i\) choose either the computational basis \(\mathcal{B}_C\) or the Fourier basis \(\mathcal{B}_F\) uniformly at random as the measurement basis. Subsequently, for each check state, all voters measure their respective particle using the same measurement basis and publish the measurement outcomes. If the computational basis is adopted and no eavesdropper exists, the measurement outcomes must form a permutation of $\{0,\,1,\,\dots,\,n-1\}$. If the Fourier basis is adopted and no eavesdropper exists, the condition \(\sum_{t=0}^{n-1} j_t \equiv 0 \pmod{n}\) must hold, where $j_{t}$ denotes the measurement outcome of voter $V_{t}$ for $t=0,\,1,\,\dots,\,n-1$. All quantum states used for detection are discarded after the detection procedure. If the detection passes, exactly \(1+nm\) unmeasured entangled states remain for the subsequent steps.

\smallskip
\noindent\hspace{1em}\refstepcounter{stepcnt}\textbf{Step \thestepcnt. Secret-index generation.}
\label{step3}
One entangled state \(|\Psi_n\rangle\) is randomly selected from the remaining \(1+nm\) states after eavesdropping detection to serve as the secret-index state. For this state, each voter measures his (or her) particle in the computational basis \(\mathcal{B}_C\). Suppose that the measurement outcome of voter \(V_i\) is \(d_i\). Owing to the computational-basis measurement correlation of \(|\Psi_n\rangle\), the tuple \((d_0, d_1, \dots, d_{n-1})\) forms a permutation of \((0, 1, \dots, n-1)\). Each voter retains their own secret index \(d_i\) and does not disclose it to any other participant. 

\smallskip
\noindent\hspace{1em}\refstepcounter{stepcnt}\textbf{Step \thestepcnt. Generation of ballot vectors.}
\label{step4}
After secret indices are generated, \(nm\) unmeasured entangled states  \(|\Psi_n\rangle\)  remain. The voters sequentially partition these $nm$ states into $n$ groups, each containing $m$ states. For ease of description, we denote the $r$-th state in the $s$-th group as \(|\Psi_n\rangle^{(s)}_{r}\), where $r\in\{1,\,2,\,\ldots,\,m\}$ and $s\in\{1,\,2,\,\ldots,\,n\}$.

 For state \(|\Psi_n\rangle^{(s)}_{r}\), voter $V_{i}$ measures the particle that he (or she) owns with the Fourier basis \(\mathcal{B}_F\) and denotes the measurement outcome as   \(a_{r,i}^{(s)}\), where $0\leq a_{r,i}^{(s)}\leq n-1$. Voter $V_{i}$ writes the measurement outcomes of the particles he owns for the states in $s$-th group in sequence as a column vector, i.e.,  
 \begin{equation}
 \nonumber
 \alpha_{i}^{(s)}=
\begin{bmatrix}
a_{0,i}^{(s)} \\
a_{1,i}^{(s)} \\
\vdots \\
a_{m-1,i}^{(s)}
\end{bmatrix},
\end{equation}
where $0\leq s\leq n-1$ and $0\leq i\leq n-1$.

According to the Fourier-basis correlation property of \(|\Psi_n\rangle\),  we have \(\sum_{i=0}^{n-1} a_{r,i}^{(s)} \equiv 0 \pmod n\) for  $r\in\{1,\,2,\,\ldots,\,m\}$ and $s\in\{1,\,2,\,\ldots,\,n\}$. 

\smallskip
\noindent\hspace{1em}\refstepcounter{stepcnt}\textbf{Step \thestepcnt. Multiple-selection ballot encoding.}
\label{step5}
Voter \(V_i\) encodes the ballot $v_{i}$ into vector $A^{(s)}_{i}$ by the following rule, i.e., 
\begin{equation}
A^{(s)}_{i}=\alpha_{i}^{(s)}+\delta_{s,d_i}v_{i}(mod\, n)
\end{equation}
where $i=0,\,1,\,\ldots,\,n-1; \,\, $\(\delta_{s,d_i}=1\) if \(s=d_i\) and \(\delta_{s,d_i}=0\) if $s\neq d_{i}$.

\smallskip
\noindent\hspace{2em}\refstepcounter{stepcnt}\textbf{Step \thestepcnt. Simultaneous broadcast and self-tallying.}
\label{step6}
Each Voter $V_{i}$ publishes his (or her) voting data, i.e., $A^{(s)}_{i}$, where $0\leq s\leq n-1$ and $0\leq i\leq n-1$. Thus, each participant can get $n$ ballot matrices, i.e., 

\[
A^{(s)} =
\begin{bmatrix}
A^{(s)}_{0} & A^{(s)}_{1} & \cdots & A^{(s)}_{n-1} \\
\end{bmatrix}.
\]
where $0\leq s\leq n-1$.

To verify the correctness of the votes, the voter $V_{i}$ computes $\beta^{(d_{i})}=\sum_{j=0}^{n-1}A^{(d_{i})}_{j}$ $(mod \,n)$, and verifies whether $\beta^{(d_{i})}=v_{i}$ holds, for $i=0,\,1,\,\ldots,\,n-1.$ If these equations hold, the voters announce to proceed with the protocol.

Each participant computes the tally vector
$$\beta=\sum_{s=0}^{n-1}\left\{\left(\sum_{i=0}^{n-1}A_{i}^{(s)}\right) (mod \,n)\right\}.$$
In fact, the $r$-th component of vector $\beta$ is the number of votes for candidate \(C_r\), where $r\in\{0,\,1,\,\ldots,n-1\}$

\subsection{An Example}

To illustrate the proposed protocol, consider an election with \(n=4\) voters, \(m=3\) candidates, and a maximum of \(L=2\) selections per voter. Thus, 13 4-particle, 4-level entangled states $\lvert\Psi_4\rangle$ are needed, of which one is used to generate secret indices and the remaining 12 are divided into 4 groups in sequence. Suppose that the secret indices are
\begin{equation}
(d_0,d_1,d_2,d_3)=(2,0,3,1),
\end{equation}
and that the submitted ballots are
\begin{equation}
\nonumber
v_{0}=
\begin{bmatrix}
1 \\
0 \\
1
\end{bmatrix},\,
v_{1}=
\begin{bmatrix}
0 \\
1 \\
0
\end{bmatrix},\,
v_{2}=
\begin{bmatrix}
1 \\
1 \\
0
\end{bmatrix},\,
v_{3}=
\begin{bmatrix}
0 \\
0 \\
1
\end{bmatrix}.
\end{equation}

It should be noted that the four particles of each of those 12 states are sent to the four voters, one particle for each voter. According to the Fourier-basis correlation property of $\lvert\Psi_4\rangle$,  the measurement outcomes satisfy 
\begin{equation}
\sum_{i=0}^{3}a_{r,i}^{(s)}\equiv0\pmod4,
\end{equation}
where $a_{r,i}^{(s)}$ denotes $V_{i}$'s measurement outcome on his (or her) $r$-th particle of the $s$-th group.

Voter \(V_i\) encodes the ballot $v_{i}$ into vector $A^{(s)}_{i}$ by the following rule, i.e., 
\begin{equation}
A^{(s)}_{i}=\alpha_{i}^{(s)}+\delta_{s,d_i}v_{i}(mod\, n)
\end{equation}
where $i=0,\,1,\,2,\,3; \,\, $\(\delta_{s,d_i}=1\) if \(s=d_i\) and \(\delta_{s,d_i}=0\) if $s\neq d_{i}$.

Each participant computes the tally vector
\begin{equation}
\nonumber
\beta=\sum_{s=0}^{3}\left[\left(\sum_{j=0}^{3}A_{j}^{(s)}\right) (mod \,n)\right]=\sum_{s=0}^{3}v_{s}=
\begin{bmatrix}
2 \\
2 \\
2
\end{bmatrix}.
\end{equation}

\section{CORRECTNESS AND SECURITY ANALYSIS}
\label{sec:correctness-security-analysis}

To specify the security model, we make the following assumptions. The trusted third party $T$ correctly prepares and distributes the quantum states and does not collude with any adversary. For each information state, \(T\) independently applies a random permutation to the particle-to-voter assignment and keeps this mapping secret for all states not selected for security checking. All classical communications are authenticated, and simultaneous broadcast is used during the vote-pubish phase. Those states used for eavesdropping detection are randomly selected only after all quantum transmissions have been completed. We assume that at least two voters are honest; equivalently, the number of dishonest voters \(t\) satisfies \(t\leq n-2\).  

\subsection{Correctness Analysis}
Now, we show that the vote tallying method is correct.
\begin{align}
	&\quad \beta\nonumber\\
	&=\sum_{s=0}^{n-1}\left\{\left(\sum_{i=0}^{n-1}A_{i}^{(s)}\right) (\bmod n)\right\} \nonumber\\
	&=\sum_{s=0}^{n-1}\left\{\left[\sum_{i=0}^{n-1}\left(\alpha_{i}^{(s)}+\delta_{s,d_i}v_{i}(\bmod n)\right)\right](\bmod n)\right\}\nonumber\\	
	&=\sum_{s=0}^{n-1}\left\{\left[\sum_{i=0}^{n-1}\alpha_{i}^{(s)}\right](\bmod n)+\left[\sum_{i=0}^{n-1}\delta_{s,d_i}v_{i}\right](\bmod n)\right\}\nonumber\\	
	&=\sum_{s=0}^{n-1}\left\{
	\begin{bmatrix}
		\sum_{i=0}^{n-1}a^{(s)}_{0,i} \\
		\sum_{i=0}^{n-1}a^{(s)}_{1,i} \\
		\vdots \\
		\sum_{i=0}^{n-1}a^{(s)}_{m-1,i}
	\end{bmatrix}
	(\bmod n)
	+\left[\sum_{i=0}^{n-1}\delta_{s,d_i}v_{i}\right](\bmod n)\right\}\nonumber\\	
	&=\sum_{i=0}^{n-1}v_{i}
\end{align}
Clearly, the $r$-th component of vector \(\boldsymbol{B}\) corresponds to the number of votes received by the $r$-th candidate.

\subsection{Privacy Analysis}
Privacy in the proposed protocol refers to the unlinkability between a recovered ballot and the identity of the voter who submitted it. After the simultaneous broadcast in Step~\ref{step6}, the ballot vectors themselves can be recovered in anonymous form. Therefore, the privacy requirement is not to hide ballot contents after tallying, but to prevent an adversary from linking any recovered ballot to its voter. Similar to the privacy analysis of the original SQAV protocol \cite{Wang2016SelfTallying}, we consider an outside eavesdropper and a coalition of dishonest voters. In the proposed protocol, voter privacy mainly relies on the secrecy of ballot vectors generated in Step~\ref{step4} and the secret indices generated in Step~\ref{step3}.
\subsubsection{Outside eavesdropper}
Suppose that an outside eavesdropper Eve intercepts particles transmitted from the trusted third party \(T\) to the voters in Step~\ref{step1}. Since the states used for eavesdropping detection are selected only after all particles have been distributed, Eve does not know in advance which states will subsequently be used as check states.

The trusted third party initially distributes \(N_{\mathrm{tot}}=1+nm+n\delta\) \(\lvert\Psi_n\rangle\). After all \(n\) voters sequentially select \(\delta\) states for security checking, \(n\delta\) states are measured and discarded, leaving \(N_{\mathrm{rem}}=1+nm\) states for secret-index and random ballot vectors generation. Suppose that Eve disturbs \(x\) distinct entangled states before the check states are selected. The probability that all \(x\) disturbed states happen to remain outside the checking set is
\begin{equation}
P_{\mathrm{esc}}=\frac{\binom{1+nm}{x}}{\binom{1+nm+n\delta}{x}}=\prod_{q=0}^{x-1}\frac{1+nm-q}{1+nm+n\delta-q}.
\label{eq:eve_escape}
\end{equation}
For fixed \(n\), \(m\), and \(x\),
\begin{equation}
P_{\mathrm{esc}}=O\left(\delta^{-x}\right),
\end{equation}
which approaches zero as the security parameter \(\delta\) increases. Thus, an attack involving a larger number of states has a smaller probability of escaping eavesdropping detection.

We next consider a disturbed state that is selected for checking. Let \(\rho_E\) denote the state received by the voters after Eve's intervention. Define \(P_C\) as the probability that the computational-basis outcomes form a permutation of \(\mathbb{Z}_n\), and \(P_F\) as the probability that the Fourier-basis outcomes satisfy \(\sum_{i=0}^{n-1}j_i\equiv0\pmod n\). Since the checker chooses the two measurement bases with equal probability, the probability that the disturbed state passes one test is
\begin{equation}
P_{\mathrm{pass}}=\frac{1}{2}P_C+\frac{1}{2}P_F.
\label{eq:single_pass}
\end{equation}
For the ideal state \(\lvert\Psi_n\rangle\), both correlation conditions are satisfied with certainty. By contrast, any nontrivial disturbance that changes these correlations gives \(P_{\mathrm{pass}}<1\). Therefore, repeated random-sampling tests suppress the probability that Eve can disturb the transmitted states without being detected.
\subsubsection{Dishonest voters}

We next consider a coalition of dishonest voters attempting to infer the private ballot vectors held by honest voters. Let \(\mathcal{D}\subset\{0,1,\ldots,n-1\}\) denote the set of dishonest voters, where \(|\mathcal{D}|=\ell\), and let \(\mathcal{H}=\{0,1,\ldots,n-1\}\setminus\mathcal{D}\) denote the set of honest voters.

For arbitrary \(s\) and \(r\), the Fourier-basis measurement outcomes satisfy
\begin{equation}
\sum_{i=0}^{n-1}a_{r,i}^{(s)}\equiv0\pmod n.
\label{eq:zerosum_privacy}
\end{equation}
The dishonest voters may share all ballot vectors that they own. They can therefore calculate
\begin{equation}
\sum_{i\in\mathcal{H}}a_{r,i}^{(s)}\equiv-\sum_{i\in\mathcal{D}}a_{r,i}^{(s)}\pmod n.
\label{eq:honest_ballot-number_sum}
\end{equation}
Thus, a coalition inevitably knows the sum of the components of the ballot vectors held by the honest voters. However, this information does not determine any individual honest voter's component of his (or her) ballot vector. This protects multiple-selection ballot encoding in Step~\ref{step5}. 

\subsubsection{Secret indices}
In Step~\ref{step3}, the secret-index \((d_0,d_1,\ldots,d_{n-1})\) is a uniformly random permutation of \(\mathbb{Z}_n\), which is generated by the local measurements of the secret-index state in the computational-basis. Each voter learns only their own measurement outcome.
Suppose that a coalition \(\mathcal{D}\) of \(\ell\) dishonest voters shares all of their measurement outcomes. Thus the coalition knows the set
\begin{equation}
\mathcal{S}_{\mathcal{D}}=\{d_i:i\in\mathcal{D}\}.
\end{equation}
Since all secret indices are distinct, it can infer the set of indices assigned to honest voters,
\begin{equation}
\mathcal{S}_{\mathcal{H}}=\mathbb{Z}_n\setminus\mathcal{S}_{\mathcal{D}}.
\label{eq:remaining_indices}
\end{equation}
However, the coalition does not know the correspondence between the elements of \(\mathcal{S}_{\mathcal{H}}\) and the individual honest voters.

For a particular honest voter \(V_i\), conditioned on the indices revealed by the dishonest voters, each remaining index is equally likely to be \(d_i\). Therefore,
\begin{equation}
\Pr\left[d_i=s\mid\{d_j:j\in\mathcal{D}\}\right]=\frac{1}{n-\ell},\qquad s\in\mathcal{S}_{\mathcal{H}}.
\label{eq:index_guess}
\end{equation}
Equivalently, the remaining \(n-\ell\) indices are randomly assigned to the \(n-\ell\) honest voters. Hence, sharing the secret indices of dishonest voters reveals only the unused index set and does not reveal the voter–index association among honest participants. 

\subsection{Other voting properties}
\subsubsection{Self-tallying}
In our protocol, any participant or other third party with access to the broadcast data can independently obtain the voting results. Specifically, for candidate \(C_r\), he (or she) calculates
$$B=\sum_{s=0}^{n-1}\left\{\left(\sum_{i=0}^{n-1}A_{i}^{(s)}\right) (mod \,n)\right\}$$
 and can obtain the vote count of any participant. Therefore, the voting results can be obtained through simple calculation without the assistance of an additional tallying authority.

\subsubsection{Nonreusability}
In the proposed multiple-selection voting protocol, nonreusability means that each eligible voter can submit only one valid ballot vector, although several candidates may be selected within that ballot. More specifically, the ballot of voter \(V_i\) must satisfy \(v_{r,i}\in\{0,1\}\) and \(\sum_{i=0}^{m-1}v_{r,i}\leq L\). Suppose that \(V_i\) attempts to cast an additional ballot by modifying another anonymous matrix indexed by \(s\neq d_i\). Since \((d_0,d_1,\ldots,d_{n-1})\) is a permutation of $\{0,1,\ldots,n-1\}$, there exists another voter \(V_j\) who owns the index number \(d_j\) such that \(d_j=s\). 

In this case, when $V_{j}$ verifies his (or her) own vote, he (or she) computes
\begin{align}\nonumber
&\quad \, \beta^{(d_{j})}\\ \nonumber
	&=\left[\sum_{t=0,t\neq i}^{n-1}A^{(d_{j})}_{t}\right]+\alpha_{i}^{(d_j)}+\delta_{d_j,d_i}v_{i}+\delta_{d_j,d_j}v_{i}(mod \,n)\\\nonumber
	&=\sum_{t=0,t\neq i}^{n-1}\left[\alpha_{t}^{(d_j)}+\delta_{d_j,d_t}v_{t}\right]+\alpha_{i}^{(d_j)}+v_{i}(mod\, n)\\ \nonumber
	&=\left[\sum_{t=0}^{n-1}\alpha_{t}^{(d_j)}\right]+\left[\sum_{t=0,t\neq i}^{n-1}\delta_{d_j,d_t}v_{t}\right]+v_{i}(mod\, n)\\ \nonumber
	&=\mathbf{0}+\delta_{d_j,d_j}v_{j}+v_{i}(mod\, n)\\ \nonumber
	&=v_{j}+v_{i}(mod\, n)\\ \nonumber
	&\neq v_{j}. \nonumber
\end{align}

This means that $V_{j}$ can detect that his (or her) voting matrix has been used for duplicate voting by others.

\subsubsection{Verifiability}
After the broadcast, each voter \(V_i\) can verify whether their ballot has been counted correctly by calculating $\beta^{(d_{i})}=\sum_{j=0}^{n-1}A^{(d_{i})}_{j}$ $(mod \,n)$ and verifying whether $\beta^{(d_{i})}=v_{i}$ holds.

On the other hand, any participant can obtain the vote count of each candidate by calculating 
$$\beta=\sum_{s=0}^{n-1}\left\{\left(\sum_{i=0}^{n-1}A_{i}^{(s)}\right) (mod \,n)\right\},$$
where the $r$-th component of vector $\beta$ is the number of votes for candidate \(C_r\) for $r\in\{0,\,1,\,\ldots,n-1\}$. Therefore, the proposed protocol provides individual ballot verifiability and public tally.

\subsubsection{Fairness}
Before the broadcast, each voter knows only their own secret index, ballot vectors, and ballot information, and cannot obtain the ballots of other voters or any partial voting result. In Step~\ref{step6}, all encoded ballot information are announced through simultaneous broadcast channel. Since each voter can only encode information based on his (or her) own secret index, and the encoded vectors are the outcomes of local measurements on states from different groups, this process is identical for every voter. Hence, the proposed protocol satisfies fairness.
\section{Comparison with Representative SQAV Protocols}
\label{subsec:comparison}

In this section, we compare our protocol with two representative SQAV protocols proposed by Wang \textit{et al.} \cite{Wang2016SelfTallying} and Yang \textit{et al.} \cite{Yang2025ResourceEfficient}. These three protocols follow the same general objective of removing a dedicated tallying authority, but they address different aspects of self-tallying quantum anonymous voting.

\begin{table*}[t]
\caption{Comparison of representative self-tallying quantum anonymous voting protocols. }
\label{tab:comparison}
\begin{ruledtabular}
\begin{tabular}{cccccc}
\textbf{Protocol} & \textbf{Ballot functionality} & \textbf{Entangled states} & \textbf{Quantum resource} & \textbf{State transmission} & \textbf{Main feature} \\
Wang et al. \cite{Wang2016SelfTallying} & \shortstack{\(v_i\in\mathbb{Z}_m\);\\single selection} & \shortstack{Multipartite\\entangled states} & \(n^{2}\log_{2}m+n\log_{2}n\) & \shortstack{Distributor\\to voters} & \shortstack{First SQAV\\framework} \\
Yang et al. \cite{Yang2025ResourceEfficient} & Single-selection voting & \shortstack{Two-particle\\entangled states} & \(mn[1+\log_{2}(n+1)]\) & \shortstack{Circular voter-to-voter\\transmission} & \shortstack{Resource-efficient\\SQAV} \\
Ours & \shortstack{\(\mathbf{v}_i\in\{0,1\}^{m}\);\\\(\sum_{r}v_{i,r}\leq L\)} & \shortstack{\(n\)-particle, \(n\)-level\\entangled states} & \(n(nm+1)\log_{2}n\) & \shortstack{ Trusted third party\\to voters} & \shortstack{Constrained multiple-\\selection SQAV} \\
\end{tabular}
\end{ruledtabular}
\end{table*}

Wang \textit{et al.} \cite{Wang2016SelfTallying} established the basic self-tallying framework for quantum anonymous voting. In their protocol, each voter submits a value \(v_i\in\mathbb{Z}_m\), which represents one candidate. Using secret index numbers, voters are anonymously mapped to ballot boxes; the published results are a random permutation of individual votes. Thus, any participant can calculate the final voting result without relying on an additional tallying authority. This construction provides an important basis for subsequent SQAV protocols. However, each ballot allows for a single candidate rather than multiple selections. Yang \textit{et al.} \cite{Yang2025ResourceEfficient} subsequently focused on reducing the quantum-resource requirement. Their protocol replaces multipartite entangled states with two-particle entangled states and reduces the quantum resource cost of the information states from \(n^{2}\log_{2}m+n\log_{2}n\) qubits in the protocol of Wang \textit{et al.} to \(mn[1+\log_{2}(n+1)]\) qubits. The protocol is also generalized to accommodate positive voting, negative voting, and abstention.  In particular, a voter still cannot simultaneously select multiple candidates in one ballot. Our protocol effectively addresses the multi-selection voting problem in quantum anonymous self-tallying. In our protocol, one voter may select several candidates within the prescribed voting rule. More importantly, the multi-selection ballot is processed as a single anonymous object rather than as several independent single-selection ballots. A concise comparison is given in Table~\ref{tab:comparison}. The quantum-resource costs listed in the table refer only to the information states and exclude the additional states used for eavesdropping detection.

Table~\ref{tab:comparison} summarizes the main differences among the three protocols in terms of ballot functionality, entangled-state structure, quantum-resource requirement, and quantum-state transmission. Wang et al. established the first SQAV framework, whereas Yang et al. substantially reduced the quantum-resource requirement by using two-particle entangled states. We propose an SQAV protocol that supports multiple-selection voting, which is not available in existing protocols.
\section{CONCLUSION}
\label{sec:conclusion}
Since the first self-tallying quantum anonymous voting protocol was proposed, it has attracted widespread interest. However, due to the difficulty of satisfying requirements such as anonymity and self-tallying, the existing SQAV protocols only support single-selection voting. Designing a SQAV protocol with multi-selection functionality is still a challenging problem that has not been effectively solved.

In this paper, we propose a SQAV protocol for multi-selection elections. In our protocol, each ballot is encoded as a constrained binary vector, which allows a voter to select multiple candidates in a single ballot. Our protocol fully exploits the correlations among different particles of those $n$-particle entangled states under Fourier-basis measurements, and effectively preserves the privacy of ballots. Our work effectively addresses the multi-selection voting problem in SAQV. 

\begin{acknowledgments}
	\vspace{-10pt}
	This work is supported by Natural Science Foundation of Shandong Province of China (Grant No. ZR2026MS1078 and ZR2023MF080), Beijing Natural Science Foundation (Grant No. 4252014) and 
Shandong Key Laboratory of Smart Mine Information Technology.
\end{acknowledgments}

\appendix

\section{Fourier-basis representation of the entangled state}
\label{app:fourier}

Throughout this appendix, the computational-basis is denoted as
\begin{equation}
	\mathcal{B}_C=\{|k\rangle_C:k=0,1,\ldots,n-1\},
\end{equation}
and the Fourier-basis is denoted as 
\begin{equation}
\mathcal{B}_F=\{|j\rangle_F:j=0,1,\ldots,n-1\}.
\end{equation}

The \(n\)-level discrete Fourier transform is defined as
\begin{equation}
|j\rangle_F=F_n|j\rangle_C=\frac{1}{\sqrt{n}}\sum_{k=0}^{n-1}\omega^{jk}|k\rangle_C,\label{eq:app_dft}
\end{equation}
where $\omega=e^{2\pi \sqrt{-1}/n}$, and \(\oplus\) denotes addition modulo \(n\).

\smallskip
\noindent\textbf{Proposition 1.}
Consider the \(n\)-particle, \(n\)-level entangled state
\begin{equation}
|\Psi_n\rangle=\frac{1}{\sqrt{n}}\sum_{k=0}^{n-1}|k,k\oplus1,\ldots,k\oplus(n-1)\rangle_C.
\label{eq:app_state}
\end{equation}
In the Fourier basis, this state can be written as
\begin{widetext}
\begin{equation}
|\Psi_n\rangle=
\frac{1}{n^{(n-1)/2}}
\sum_{\substack{j_0,\ldots,j_{n-1}\in\mathbb{Z}_n\\
\sum_{\ell=0}^{n-1}j_\ell\equiv0\;(\mathrm{mod}\;n)}}
\omega^{-\sum_{\ell=0}^{n-1}\ell j_\ell}
|j_0,j_1,\ldots,j_{n-1}\rangle_F.
\label{eq:app_fourier_rep}
\end{equation}
\end{widetext}

\smallskip
\noindent\textit{Proof.}
By Eq.~\eqref{eq:app_dft} and the unitarity of the discrete Fourier transform, we have \(F_n^{-1}=F_n^\dagger\). Thus, a computational-basis state can be expressed in the Fourier basis as
\begin{equation}
|k\rangle_C=\frac{1}{\sqrt{n}}\sum_{j=0}^{n-1}\omega^{-jk}|j\rangle_F.
\label{eq:app_inverse_dft}
\end{equation}
For the computational-basis state \(|k\oplus\ell\rangle_C\),  we have
\begin{equation}
|k\oplus\ell\rangle_C=
\frac{1}{\sqrt{n}}\sum_{j_\ell=0}^{n-1}
\omega^{-j_\ell(k\oplus\ell)}|j_\ell\rangle_F.
\label{eq:app_particle_expand}
\end{equation}
by Eq.~\eqref{eq:app_inverse_dft}, where $0 \leqslant \ell\leqslant n-1$.
Since \(\oplus\) denotes addition modulo \(n\), there exists an integer \(q\) such that
\begin{equation}
k\oplus\ell=k+\ell-qn.
\end{equation}
Because \(\omega^n=1\), the phase factor in Eq.~\eqref{eq:app_particle_expand} satisfies
\begin{align}
\omega^{-j_\ell(k\oplus\ell)}
&=\omega^{-j_\ell(k+\ell-qn)} \\
&=\omega^{-j_\ell(k+\ell)}\omega^{j_\ell qn} \\
&=\omega^{-j_\ell(k+\ell)}.
\end{align}
Hence,
\begin{equation}
|k\oplus\ell\rangle_C=
\frac{1}{\sqrt{n}}\sum_{j_\ell=0}^{n-1}
\omega^{-j_\ell(k+\ell)}|j_\ell\rangle_F.
\label{eq:app_particle_simplified}
\end{equation}

For a fixed \(k\), the \(n\)-particle state in Eq.~\eqref{eq:app_state} can be written as
\begin{equation}
|k,k\oplus1,\ldots,k\oplus(n-1)\rangle_C
=
\bigotimes_{\ell=0}^{n-1}|k\oplus\ell\rangle_C.
\end{equation}
Expanding the tensor product by using Eq.~\eqref{eq:app_particle_simplified}, we have
\begin{equation}
\begin{aligned}
&|k,k\oplus1,\ldots,k\oplus(n-1)\rangle_C\\
=&
\bigotimes_{\ell=0}^{n-1}
\left(
\frac{1}{\sqrt{n}}
\sum_{j_\ell=0}^{n-1}
\omega^{-j_\ell(k+\ell)}
|j_\ell\rangle_F
\right).
\end{aligned}
\end{equation}
Since there are \(n\) particles, the normalization factor is
\begin{equation}
\left(\frac{1}{\sqrt{n}}\right)^n=\frac{1}{n^{n/2}}.
\end{equation}
Expanding the tensor product gives
\begin{widetext}
\begin{equation}
|k,k\oplus1,\ldots,k\oplus(n-1)\rangle_C
=
\frac{1}{n^{n/2}}
\sum_{j_0,\ldots,j_{n-1}\in\mathbb{Z}_n}
\left(
\prod_{\ell=0}^{n-1}
\omega^{-j_\ell(k+\ell)}
\right)
|j_0,j_1,\ldots,j_{n-1}\rangle_F.
\label{eq:app_tensor_expand}
\end{equation}
By
\begin{equation}
\prod_{\ell=0}^{n-1}\omega^{-j_\ell(k+\ell)} = \omega^{-\sum_{\ell=0}^{n-1}j_\ell(k+\ell)},
\end{equation}
we have
\begin{equation}
|k,k\oplus1,\ldots,k\oplus(n-1)\rangle_C = \frac{1}{n^{n/2}} \sum_{j_0,\ldots,j_{n-1}\in\mathbb{Z}_n} \omega^{-\sum_{\ell=0}^{n-1}j_\ell(k+\ell)} |j_0,j_1,\ldots,j_{n-1}\rangle_F.
\label{eq:app_fixed_k}
\end{equation}

By Eq.~\eqref{eq:app_state} and Eq.~\eqref{eq:app_fixed_k}, we have

\begin{align}
|\Psi_n\rangle
&=
\frac{1}{\sqrt{n}}
\sum_{k=0}^{n-1}
\frac{1}{n^{n/2}}
\sum_{j_0,\ldots,j_{n-1}\in\mathbb{Z}_n}
\omega^{-\sum_{\ell=0}^{n-1}j_\ell(k+\ell)}
|j_0,\ldots,j_{n-1}\rangle_F \notag\\
&=
\frac{1}{n^{(n+1)/2}}
\sum_{j_0,\ldots,j_{n-1}\in\mathbb{Z}_n}
\left(
\sum_{k=0}^{n-1}
\omega^{-\sum_{\ell=0}^{n-1}j_\ell(k+\ell)}
\right)
|j_0,\ldots,j_{n-1}\rangle_F.
\label{eq:app_substitute}
\end{align}

The exponent in Eq.~\eqref{eq:app_substitute} can be rewritten as
\begin{align}
\sum_{\ell=0}^{n-1}j_\ell(k+\ell)
&=
\sum_{\ell=0}^{n-1}kj_\ell
+
\sum_{\ell=0}^{n-1}\ell j_\ell \\
&=
k\sum_{\ell=0}^{n-1}j_\ell
+
\sum_{\ell=0}^{n-1}\ell j_\ell.
\end{align}
Consequently,
\begin{equation}
\omega^{-\sum_{\ell=0}^{n-1}j_\ell(k+\ell)}
=
\omega^{-k\sum_{\ell=0}^{n-1}j_\ell}
\omega^{-\sum_{\ell=0}^{n-1}\ell j_\ell}.
\end{equation}
Using this relation, Eq.~\eqref{eq:app_substitute} can be rewritten as

\begin{equation}
|\Psi_n\rangle=
\frac{1}{n^{(n+1)/2}}
\sum_{j_0,\ldots,j_{n-1}\in\mathbb{Z}_n}
\omega^{-\sum_{\ell=0}^{n-1}\ell j_\ell}
\left(
\sum_{k=0}^{n-1}
\omega^{-k\sum_{\ell=0}^{n-1}j_\ell}
\right)
|j_0,j_1,\ldots,j_{n-1}\rangle_F.
\label{eq:app_inner_sum}
\end{equation}
\end{widetext}

Let
\begin{equation}
S=\sum_{\ell=0}^{n-1}j_\ell.
\end{equation}
The inner sum in Eq.~\eqref{eq:app_inner_sum} is
\begin{equation}
\sum_{k=0}^{n-1}\omega^{-kS}.
\end{equation}
If \(S\equiv0\pmod n\), then \(\omega^{-S}=1\), and hence
\begin{equation}
\sum_{k=0}^{n-1}\omega^{-kS}
=
\sum_{k=0}^{n-1}1
=
n.
\end{equation}
If \(S\not\equiv0\pmod n\), then \(\omega^{-S}\neq1\). Using the finite geometric-series formula, we have
\begin{align}
\sum_{k=0}^{n-1}\omega^{-kS}
&=
\frac{1-(\omega^{-S})^n}{1-\omega^{-S}} \\
&=
\frac{1-\omega^{-nS}}{1-\omega^{-S}} \\
&=0,
\end{align}
where the last equality follows from \(\omega^n=1\). Therefore,
\begin{equation}
\sum_{k=0}^{n-1}\omega^{-kS}
=
\begin{cases}
n, & S\equiv0\pmod n,\\
0, & S\not\equiv0\pmod n.
\end{cases}
\label{eq:app_roots}
\end{equation}
 Therefore,
\begin{widetext}
\begin{align}
|\Psi_n\rangle
&=
\frac{n}{n^{(n+1)/2}}
\sum_{\substack{j_0,\ldots,j_{n-1}\in\mathbb{Z}_n\\
\sum_{\ell=0}^{n-1}j_\ell\equiv0\;(\mathrm{mod}\;n)}}
\omega^{-\sum_{\ell=0}^{n-1}\ell j_\ell}
|j_0,j_1,\ldots,j_{n-1}\rangle_F \notag\\
&=
\frac{1}{n^{(n-1)/2}}
\sum_{\substack{j_0,\ldots,j_{n-1}\in\mathbb{Z}_n\\
\sum_{\ell=0}^{n-1}j_\ell\equiv0\;(\mathrm{mod}\;n)}}
\omega^{-\sum_{\ell=0}^{n-1}\ell j_\ell}
|j_0,j_1,\ldots,j_{n-1}\rangle_F.
\end{align}
\end{widetext}

% The \nocite command causes all entries in a bibliography to be printed out
% whether or not they are actually referenced in the text. This is appropriate
% for the sample file to show the different styles of references, but authors
% most likely will not want to use it.
\nocite{*}

\bibliography{Reference}% Produces the bibliography via BibTeX.

\end{document}